\documentstyle[emulateapj]{article}

\lefthead{Nagao et al.}
\righthead{NGC 4151}

\begin{document}

\submitted{The Astronomical Journal, in Press} 
\title{Subaru High-Dispersion Spectroscopy of Narrow-Line Region \\
       in the Seyfert Galaxy NGC 4151\footnote{
         Based on data collected at Subaru Telescope, which is operated
         by the National Astronomical Observatory of Japan.}
       }

\author{Tohru NAGAO, Takashi MURAYAMA, Yasuhiro SHIOYA, and Yoshiaki TANIGUCHI}
\affil{Astronomical Institute, Graduate School of Science, 
       Tohoku University, Aramaki, Aoba, Sendai 980-8578, Japan\\
       tohru@astr.tohoku.ac.jp, murayama@astr.tohoku.ac.jp, 
       shioya@astr.tohoku.ac.jp, tani@astr.tohoku.ac.jp}

%-----------------------------------------------------------------------------

\begin{abstract}

We report on a study of forbidden emission-line spectrum of 
nearby Seyfert 1.5 galaxy NGC 4151 based on the high-resolution ($R \sim
45,000$) optical spectrum obtained by using the High Dispersion Spectrograph
boarded on the Subaru Telescope.
The profile parameters such as the emission-line widths, the velocity shifts
from the recession velocity of the host galaxy, and the asymmetry indices,
for emission lines including very faint ones such as 
[Ar {\sc iv}]$\lambda \lambda$4712,4740 and
[Fe {\sc vi}]$\lambda \lambda$5631,5677, are investigated.
Statistically significant correlations between the measured profile 
parameters and the critical densities of transitions are found while
there are no meaningful correlations between the profile parameters 
and the ionization potentials of ions. 
By comparing the results with photoionization model calculations,
we remark that a simple power-law distribution of the
gas density which is independent of the radius from the nucleus cannot
explain the observed correlation between the emission-line widths and
the critical densities of the transitions.
Taking the additional dense gas component expected to exist at
the innermost of the narrow-line region into account, the observed 
correlations between the emission-line width and the critical density
of the transitions can be understood since high-critical-density
emission lines can arise at such relatively inner regions even if
their ionization potentials are low.
The observed correlation between the blueshift amounts of emission lines
and the critical densities of the ions is also explained if such 
dense gas clouds located closer 
to the nucleus have larger outflowing velocities.

\end{abstract}

\keywords{
galaxies: active {\em -}
galaxies: individual (NGC 4151) {\em -}
galaxies: ISM {\em -}
galaxies: nuclei {\em -}
galaxies: Seyfert}

%-----------------------------------------------------------------------------

\section{INTRODUCTION}

\begin{figure*}
%\epsscale{1.80}
%\plotone{Nagao.fig01.eps}
\vspace{104mm}
\caption{
Flux-calibrated, blaze-profile-corrected spectra of
the nucleus of NGC 4151 with the aperture size of 1.44 arcsec.
The sky emission is not subtracted.
The spectrum covered by the blue CCD which include the echelle order
from the 108th to the 141th is presented in the upper panel.
The spectrum covered by the red CCD which include the echelle order
from the 86th to the 106th is presented in the lower panel.
\label{fig1}}
\end{figure*}

The narrow-line region (NLR) is one of fundamental ingredients of
active galactic nuclei (AGNs), and many efforts have been made up to now
to understand its physical, chemical, geometrical, and kinematic 
properties (see, e.g., Osterbrock \& Mathews 1986). 
High spatial-resolution imaging observations have shown that NLRs 
consist of inhomogeneous and discrete gas 
clouds which are sometimes confined in a biconical structure
(e.g., Tadhunter \& Tsvetanov 1989; Evans et al. 1993; 
Macchetto et al. 1994; Boksenberg et al. 1995; 
Capetti, Axon, \& Macchetto 1997; Ferruit et al. 1999). 
Some spectroscopic observations, on the other hand, have suggested that
there is a density and/or ionization stratification in NLRs, which has
been clarified mainly by correlations between emission-line widths and
their ionization potential and/or critical density (e.g., 
Filippenko \& Halpern 1984; De Robertis \& Osterbrock 1984, 1986; 
Filippenko 1985; Veilleux 1991a; Alloin et al. 1992;
see also Barth et al. 1999; Tran, Cohen, \& Villar-Martin 2000).
Thus it is now recognized that there are many discrete gas clouds 
with various properties, which are stratified in NLRs (see, e.g., 
Ferguson et al. 1997a; Ferguson, Korista, \& Ferland 1997b).

This naturally leads to the following question; how does the density and/or 
ionization stratification exist in NLRs? In order to investigate this issue,
high spatial-resolution spectroscopic observations have been performed
recently (e.g., Kraemer et al. 2000; Kraemer \& Crenshaw 2000).
However, there are two drawbacks in this approach.
First, this method is applicable only to very nearby AGNs.
Each gas cloud in NLRs of rather distant AGNs is hard to be resolved 
spatially even by using HST. 
And second, even as for very nearby AGNs, the innermost
region of NLRs (i.e., $\lesssim$ 1 pc) cannot be resolved spatially
by the current ultraviolet/optical/infrared facilities.
Since highly ionized gas clouds may be located in such an innermost region
(e.g., Murayama \& Taniguchi 1998a, 1998b; 
Nagao, Taniguchi, \& Murayama 2000; 
Nagao, Murayama, \& Taniguchi 2001b, 2001c; see also
Barth et al. 1999; Tran et al. 2000),
this drawback seems to be a very serious problem for studies of the 
structure of ionization and physical properties of NLRs.

One of complementary approaches for the high spatial-resolution observations
is high spectral-resolution observations. By using high-dispersion
spectra, we can investigate emission-line flux ratios as a function of
a recession velocity, which contains information about the kinematic
configuration of ionized gas clouds with various physical properties.
Although high-dispersion spectra of only some very strong emission lines
such as [O {\sc iii}]$\lambda$5007 were taken to investigate 
kinematics of NLR clouds (e.g., Glaspey et al. 1976;
Pelat \& Alloin 1980; Pelat, Alloin, \& Fosbury 1981; 
Pelat \& Alloin 1982; Alloin et al. 1983; Vrtilek \& Carleton 1985),
it is difficult to explore the whole physical properties of NLR clouds
solely from the [O {\sc iii}]$\lambda$5007 line.
In order to explore the nature of NLRs from the kinematical viewpoint,
medium-resolution spectra of multi emission lines have been also 
investigated (e.g., De Robertis \& Osterbrock 1984, 1986).
However, Veilleux (1991c) reported
that such medium-resolution spectroscopy tends to overestimate the
emission-line widths while to underestimate the line asymmetries,
owing to the instrumental broadening effect.
Multi emission-line spectra with both high spectral resolution and 
high S/N are thus crucially necessary to understand the NLR clouds.

Appenzeller \& \"{O}streicher (1988) examined profiles of
forbidden emission lines including some high-ionization emission lines 
by using high-resolution spectra ($\Delta v \sim 15$ km s$^{-1}$).
However, the signal-to-noise ratios of their data
are not high enough to study the velocity profiles of weak lines in detail.
Veilleux (1991b) presented multi emission-line spectra with both high 
spectral resolution ($\Delta v \sim$ 10 km s$^{-1}$) and high S/N
for 16 bright Seyfert galaxies (see also Veilleux 1991a, 1991c).
However, the main concern of this work was to explore the statistical
properties of NLR kinematics in general and thus the profiles of
weak but important emission lines should be investigated further.
Thus we have started a program in which optical spectra with a very high
spectral resolution ($\Delta v \sim 7$ km s$^{-1}$) and a very high 
signal-to-noise ratio for some nearby Seyfert galaxies are collected 
by using high-dispersion spectrograph (HDS; Noguchi et al. 2002) 
boarded on the 8.2m Subaru telescope (Kaifu 1998).
In this paper, we report the first result of this program on the
nearby Seyfert 1.5 galaxy NGC 4151\footnote{
Adopting a distance to NGC 4151 of 13.3 Mpc, 0$\farcs$1 corresponds to a 
linear scale of 6.4 pc in the plane of the sky.}
(see Ulrich 2000 for a summary of the past observations for NGC 4151;
see also Kaiser et al. 2000, Nelson et al. 2000 and Kraemer et al. 2000 
for the recent results of a high spatial resolution spectroscopy 
of this object).

\section{OBSERVATION AND DATA REDUCTION}

%------------------------------------------------------------------
%                Table 1
%------------------------------------------------------------------
\begin{deluxetable}{lccc}
%\scriptsize
\tablenum{1}
\tablecaption{Rest Wavelengths, Critical Densities, and Ionization Potentials
              of Forbidden Emission Lines Studied in This Paper}
\tablewidth{360pt}
\tablehead{
\colhead{Emission Line} &
\colhead{$\lambda$\tablenotemark{a}} &
\colhead{Critical Density} &
\colhead{Ionization Potential} \\
\colhead{} &
\colhead{(${\rm \AA}$)} &
\colhead{(cm$^{-3}$)} &
\colhead{(eV)} 
}
\startdata  
[O {\sc iii}]\dotfill  & 4363.21 & $3.3\times10^7$ &  35.1  \nl
[Ar {\sc iv}]\dotfill  & 4711.34 & $1.3\times10^4$ &  40.7 \nl
[Ar {\sc iv}]\dotfill  & 4740.20 & $1.2\times10^5$ &  40.7 \nl
[O {\sc iii}]\dotfill  & 4958.92 & $7.0\times10^5$ &  35.1 \nl
[O {\sc iii}]\dotfill  & 5006.85 & $7.0\times10^5$ &  35.1 \nl
[Fe {\sc vi}]\dotfill  & 5630.82 & \nodata         &  75.0 \nl
[Fe {\sc vi}]\dotfill  & 5676.96 & \nodata         &  75.0 \nl
[Fe {\sc vii}]\dotfill & 5721.11 & $3.6\times10^7$ &  99.1 \nl
[N {\sc ii}]\dotfill   & 5754.57 & $3.2\times10^7$ &  14.5 \nl
[Fe {\sc vii}]\dotfill & 6086.92 & $3.6\times10^7$ &  99.1 \nl
[O {\sc i}]\dotfill    & 6300.32 & $1.8\times10^6$ &   0.0 
\enddata 
\tablenotetext{a}{Taken from Bowen (1960).}
\end{deluxetable}
%------------------------------------------------------------------

 \subsection{Observation}

We carried out spectroscopic observation of the Seyfert galaxy NGC 4151
by using HDS boarded on the Subaru telescope, on the 4th February 2002 (UT).
The detector of HDS is a mosaic of two 4k $\times$ 2k EEV CCD's with
13.5 $\mu$m pixels. We used the grating for the red spectra, whose
central wavelength is set to 5600 ${\rm \AA}$.
By this setting, one CCD covers a wavelength range of 
$4250 {\rm \AA} - 5550 {\rm \AA}$ (from the 108th order to the 141st order)
and the other CCD covers a range of $5650 {\rm \AA} - 6950 {\rm \AA}$
(from the 86th order to the 106th order).
A short-wavelength-cutoff filter of KV389 was used to obtain the whole
wavelength range written above.
We adopted 2 pixel binning for both wavelength and spatial directions 
on the chips. As a result of this binning, the spatial sampling rate was
0.24 arcsec/binnedpixel.
By adopting a 0.80 arcsec (0.40 mm)
slit width, we got spectra with a wavelength resolution of $\sim$45,000.
The atmospheric dispersion compensator (ADC) was used in order to avoid
photon losses caused by the atmospheric dispersion effect.
The image rotator was also used by adopting the red setup. Note that 
the ADC and the image rotator do not degrade the spatial resolution.
See Noguchi et al. (2002) for the wavelength dependences of the throughput of
the ADC and the image rotator.

We carried out four 900 sec exposures for NGC 4151 with the position
angle of 77$\arcdeg$
The typical seeing size was $\sim 1\farcs5$ during the observation.
We also obtained the spectrum of a spectroscopic standard star,
BD+26$\arcdeg$ 2606 (Oke \& Gunn 1983), for flux calibration.
The spectra of a halogen lamp and a thorium-argon lamp were also obtained
for the flat field and the wavelength calibration, respectively.

 \subsection{Data Reduction}

The data were reduced with the IRAF\footnote{
IRAF (Image Reduction and Analysis Facility) is distributed by the 
National Optical Astronomy Observatories, which is operated by the
Association of Universities for Research in Astronomy, Inc. under
corporative agreement with the National Science Foundation.
} package in the standard manner. Since the expected dark current of the
CCD's is $\sim$1 ADU/pixel for an hour (see Noguchi et al. 2002), 
we neglected the contribution of the dark current in this study.
Although the data frames obtained by HDS include over-scan regions to be
used for the bias subtraction, we did not use the over-scan data for
the bias subtraction because the time variation of the bias level 
estimated by using the over-scan data was negligibly small.
Instead of using the over-scan data, we simply subtracted the bias frame
taken as the calibration data from the object frames.
The cosmic-ray noise was removed by running the IRAF task ``lineclean''
for the object frames divided by the median-filtered object frames.
The flat fielding was performed by dividing the object frames by the
halogen lamp frame directly. By this process, the blaze profiles of the
object frames were corrected very roughly (see Section 2.4 and Appendix A).
The background subtraction was done by the ``apscatter'' routine of IRAF.
This background component is thought to be attributed to reflections
of the object light at the surface of the CCD's and the field flattener
lens just before the CCD's (see Noguchi et al. 2002 for details).
The wavelength calibration was carried out by using the thorium-argon 
spectra obtained before and after the observations of the objects.
The measured rms wavelength error of the thorium-argon lines is
less than 0.005 ${\rm \AA}$, which corresponds to 0.2 pixels at
the bluest region in the spectra. By measuring the FWHM of thorium-argon 
lines in the wavelength-calibrated thorium-argon spectra, we confirmed 
the spectral resolution to be $\gtrsim 45,000$. 

We extracted the nuclear spectra of the objects
adopting the aperture size of 1.44 arcsec (i.e., 6 binned pixels),
by using the IRAF task ``apall''. Here we should mention that the 
resultant aperture for the extraction of the nuclear spectrum is 
$0.80^{\prime \prime} \times 1.44^{\prime \prime}$, which is
rather smaller than the spatial extension of the ionized gas clouds 
around the nucleus of NGC 4151 (e.g., Heckman \& Balick 1983;
Unger et al. 1987; Hutchings et al. 1998, 1999; Winge et al. 1999; 
see also Kaiser et al. 2000; Nelson et al. 2000; Kraemer et al. 2000). 
Although a large part of the NLR emission of NGC 4151 is covered by 
this adopted aperture since the position  angle of the slit is along 
to the direction of the NLR extension (77 deg), some ionized gas clouds 
are missed in our analysis. We will see how the possible aperture 
effects can affect our discussion when necessary (in Section 4).

\begin{figure*}
\epsscale{1.60}
\plotone{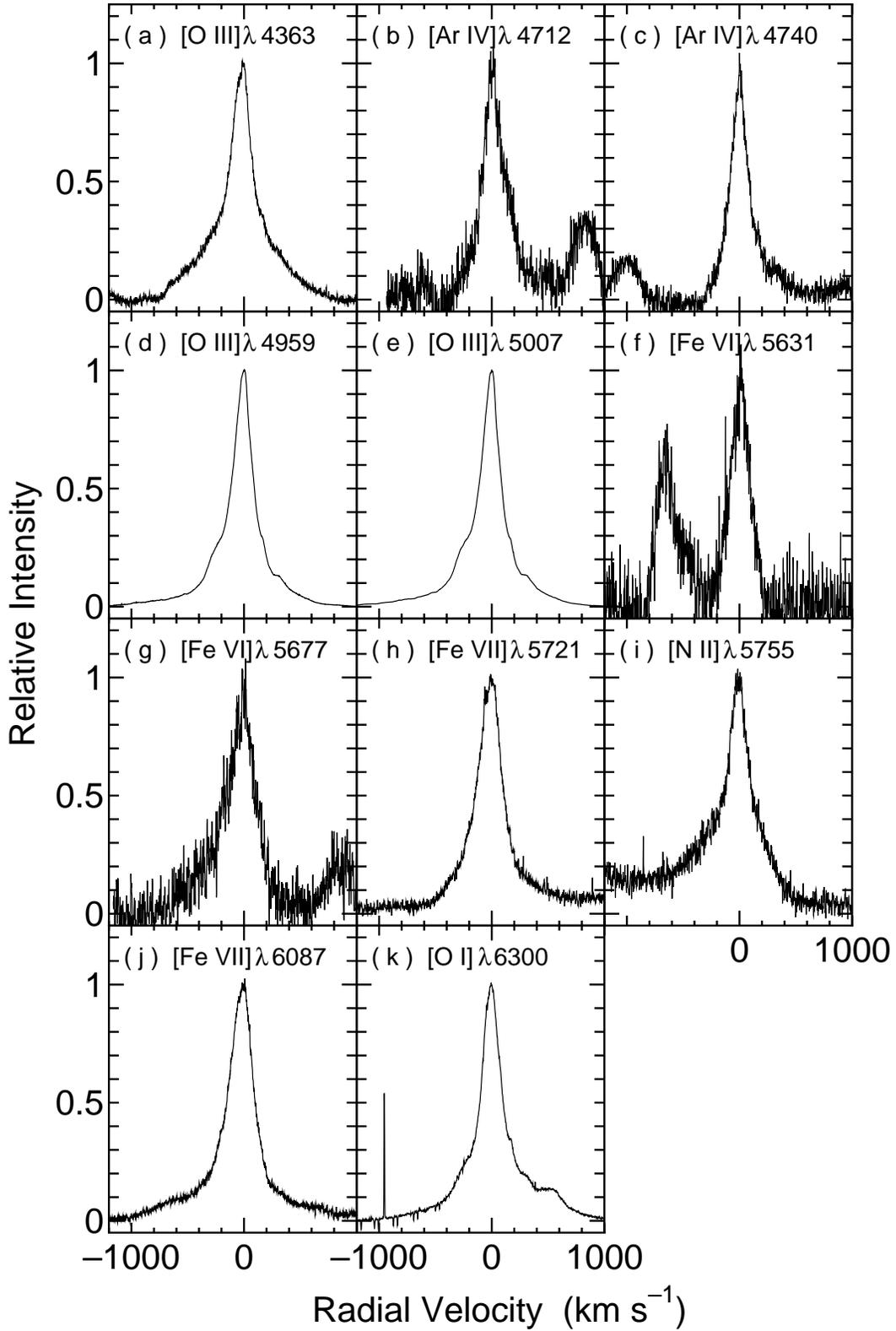}
\caption{
Profiles of the continuum-subtracted forbidden emission lines studied
in this paper.
The origin of the horizontal axis coincides with the recession velocity
of the peak of the [O {\sc iii}]$\lambda$5007 emission, which is
located at 5022.82 ${\rm \AA}$ in the obtained spectrum.
The spectral profile in the relative recession-velocity range of 
--1200 km s$^{-1} \leq \Delta v \leq +1000$ km s$^{-1}$ is shown
for each emission line.
The flux level of each emission line is normalized
by the flux density at the peak of the line.
The instrumental broadening effect is not corrected.
(a) [O {\sc iii}]$\lambda$4363,
(b) [Ar {\sc iv}]$\lambda$4712,
(c) [Ar {\sc iv}]$\lambda$4740,
(d) [O {\sc iii}]$\lambda$4959,
(e) [O {\sc iii}]$\lambda$5007,
(f) [Fe {\sc vi}]$\lambda$5631,
(g) [Fe {\sc vi}]$\lambda$5677,
(h) [Fe {\sc vii}]$\lambda$5721,
(i) [N {\sc ii}]$\lambda$5755,
(j) [Fe {\sc vii}]$\lambda$6087, and
(k) [O {\sc i}]$\lambda$6300.
\label{fig2}}
\end{figure*}

\begin{figure*}
\epsscale{1.40}
\plotone{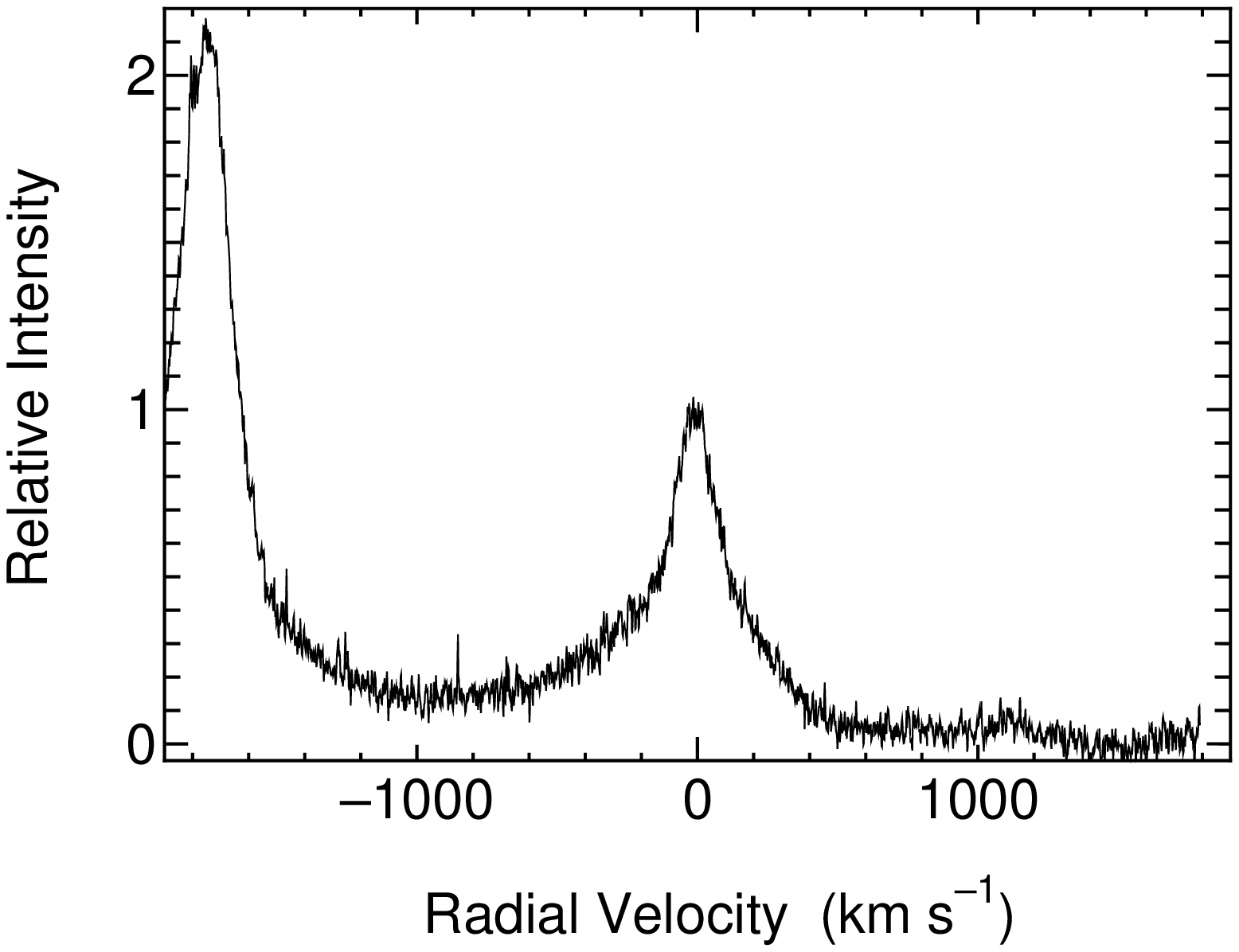}
\caption{
Continuum-subtracted spectrum of NGC 4151 around the 
[N {\sc ii}]$\lambda$5755 emission. The flux level is normalized by the
peak flux density of the [N {\sc ii}]$\lambda$5755 emission. 
\label{fig3}}
\end{figure*}

 \subsection{Flux Calibration}

We performed the flux calibration by using the spectrum of a 
spectroscopic standard star, BD+26$\arcdeg$ 2606, in order to correct
the wavelength dependence of the sensitivity.
Note that the absolute flux scale is not so accurate since
the seeing size was so large that the slit loss of the photons
is not negligible. This ambiguity of the flux scale does not affect
the following analysis and discussion, because we do not refer any
absolute fluxes of the emission lines and the continuum emission
in this study.

In order to estimate the sensitivity function for each echelle order,
the data of the absolute spectral energy distribution (SED) of
BD+26$\arcdeg$ 2606, sampled by a few ${\rm \AA}$, are necessary.
We made the SED data by performing a 3rd polynomial fitting for
the data presented by Oke \& Gunn (1983).
Since BD+26$\arcdeg$ 2606 is a typical A-type star ($m_V = 9.7$),
we can estimate the absolute SED by this fitting, except for the
wavelength regions where the absorption feature is significant
(e.g., the wavelength regions around H$\alpha$ and H$\beta$).
We then determined the sensitivity function for each echelle order 
by comparing the observed spectra (in ADU) with the estimated
absolute SED. As for the wavelength regions where the absolute SED
cannot be estimated due to absorption features, we estimated
the sensitivity functions by interpolating those of the neighboring 
echelle orders.
The flux calibration of the spectra of NGC 4151 was then performed by
using the estimated sensitivity functions.
Note that the terrestrial absorption features are not corrected
in this flux calibration manner. Such absorption features do not
affect the analysis and the discussion.

 \subsection{Correction for the Blaze Profile}

In general, spectra taken by echelle spectrographs have a blaze
profile similar to a sinc function, which should be corrected
in order to investigate some spectral features. The blaze profile
can be in principle corrected by dividing the object frames by
the flat frame, if the blaze profile would not change at all.
However, the blaze profile of HDS is currently unstable with no 
systematic trend, i.e., the blaze profiles are different among the 
object frames, the standard-star frames, and the flat frames (see 
Aoki 2002). Due to this unstability of the blaze profile, significant
artificial periodic features appear on the flux-calibrated
order-combined spectrum. Since the amplitude of such features reaches
up to $\sim 20$ \% of the spectrum itself, it should be corrected in
order to analyze the profiles of emission lines appropriately. 
Therefore we have carried out the blaze profile correction by 
adopting a method described at Appendix A in detail.
As a result of the correction, the amplitude of the remaining
artificial features is $< 5$ \% of the continuum level.

\section{RESULTS}

%------------------------------------------------------------------
%                Table 2
%------------------------------------------------------------------
\begin{deluxetable}{lccccccc}
%\scriptsize
\tablenum{2}
\tablecaption{Measured Profile Parameters for Observed Emission Lines}
%\tablewidth{360pt}
\tablehead{
\colhead{Emission Line} &
\colhead{$W80$} &
\colhead{$W50$} &
\colhead{$W20$} &
\colhead{$C80$} &
\colhead{$C50$} &
\colhead{$C20$} &
\colhead{$AI_{20}$\tablenotemark{a}} \\
\colhead{} &
\colhead{(km/s)} &
\colhead{(km/s)} &
\colhead{(km/s)} &
\colhead{(km/s)} &
\colhead{(km/s)} &
\colhead{(km/s)} &
\colhead{}
}
\startdata  
[O {\sc iii}]$\lambda$4363\dotfill & 
   115 & 240 & 690 & --49 & --54 & --62 & +0.04 \nl
[Ar {\sc iv}]$\lambda$4712\dotfill & 
    87 & 223 & 430 & --24 &  --9 &$\pm$0&--0.11 \nl
[Ar {\sc iv}]$\lambda$4740\dotfill & 
    64 & 177 & 365 & --27 & --26 &  --8 &--0.11 \nl
[O {\sc iii}]$\lambda$4959\dotfill & 
    94 & 211 & 504 & --37 & --40 & --74 & +0.15 \nl
[O {\sc iii}]$\lambda$5007\dotfill & 
    94 & 210 & 513 & --36 & --40 & --80 & +0.17 \nl
[Fe {\sc vi}]$\lambda$5631\dotfill & 
    72 & 191 & 354 & --15 & --23 & --28 & +0.08 \nl
[Fe {\sc vi}]$\lambda$5677\dotfill & 
    82 & 278 & 574 & --32 & --60 &--106 & +0.26 \nl
[Fe {\sc vii}]$\lambda$5721\dotfill& 
   128 & 250 & 522 & --38 & --41 & --61 & +0.09 \nl
[N {\sc ii}]$\lambda$5755\dotfill  & 
   109 & 258 &\nodata\tablenotemark{b}& 
   --40 & --42 &\nodata\tablenotemark{b}&\nodata\tablenotemark{b}\nl
[Fe {\sc vii}]$\lambda$6087\dotfill& 
   132 & 254 & 497 & --45 & --55 & --75 & +0.12 \nl
[O {\sc i}]$\lambda$6300\dotfill   & 
   106 & 206 & 565 & --36 & --29 & --34 &--0.01
\enddata 
\tablenotetext{a}{Asymmetry parameter, $AI_{20} = (C80 - C20) / W20$.
                  See Heckman et al. (1981).}
\tablenotetext{b}{Not measured since the blue wing of the 
                  [N {\sc ii}]$\lambda$5755 emission is significantly 
                  affected by the red tail of the
                  [Fe {\sc vii}]$\lambda$5721 emission.}
\end{deluxetable}
%------------------------------------------------------------------

We present the flux-calibrated, blaze-profile-corrected spectrum of
NGC 4151 in Figure 1. Note that the sky emission is not subtracted
and thus the terrestrial [O {\sc i}]$\lambda \lambda$6300,6363 
airglow emission lines are also seen in this spectrum. 
These sky emission lines can be used to check the accuracy of the
wavelength calibration. The measured wavelengths of these two emission
lines are 6300.30 ${\rm \AA}$ and 6363.78 ${\rm \AA}$, which are
consistent with the theoretical wavelengths within 
$\simeq$ 0.01 ${\rm \AA}$.

We show the profiles of the continuum-subtracted forbidden emission lines,
i.e., [O {\sc iii}]$\lambda$4363, [Ar {\sc iv}]$\lambda$4712,
[Ar {\sc iv}]$\lambda$4740, [O {\sc iii}]$\lambda$4959, 
[O {\sc iii}]$\lambda$5007, [Fe {\sc vi}]$\lambda$5631, 
[Fe {\sc vi}]$\lambda$5677, [Fe {\sc vii}]$\lambda$5721,
[N {\sc ii}]$\lambda$5755, [Fe {\sc vii}]$\lambda$6087, and
[O {\sc i}]$\lambda$6300, in Figures 2a -- 2k. 
The presented profiles are not corrected for the instrumental broadening
since the effects of the instrumental broadening are evidently negligible.
Indeed the [O {\sc i}]$\lambda$6300 airglow emission seen in Figure 2k
is much significantly narrower than the velocity features of the observed
forbidden emission lines of NGC 4151.
The origin of the horizontal axis coincides with the recession velocity
of the peak of the [O {\sc iii}]$\lambda$5007 emission, which is
located at 5022.82 ${\rm \AA}$ in the obtained spectrum.
See Table 1 for the rest wavelengths we adopted for the forbidden lines,
which are taken from Bowen (1960).
The critical density of the transition and the ionization potential
of the corresponding ion for each emission line are also given in this table.
The spectral profile in the relative recession-velocity range of 
--1200 km s$^{-1} \leq \Delta v \leq +1000$ km s$^{-1}$ is shown
for each emission line.
The continuum level for each emission line
is estimated by interpolating the neighbor wavelength regions
of the emission line linearly, except for [O {\sc iii}]$\lambda$4363.
Since this emission line is located on the broad wing of the H$\gamma$ 
emission, we extract the profile of [O {\sc iii}]$\lambda$4363
by interpolating the neighbor wavelength regions adopting
the cubic spline fitting.
In this figure, the flux level of each emission line is normalized
by the flux density at the peak of the line, which is measured
after performing a 5 pixels smoothing for the spectra.
Note that we do not make attempts to extract
the velocity profiles of blending forbidden lines, e.g., 
[O {\sc i}]$\lambda$6363, [Fe {\sc x}]$\lambda$6374,
[N {\sc ii}]$\lambda \lambda$6548,6583, and 
[S {\sc ii}]$\lambda \lambda$6717,6731. This is because 
we cannot deblend the blended lines without some assumptions
whose validity is hard to be examined.
Because of the same reason, we do not investigate velocity profiles of
the narrow component of permitted lines.

Here we should mention the contributions of other weak emission lines
to the spectra presented in Figure 2.
The emission line seen both at the red side of [Ar {\sc iv}]$\lambda$4712 and
at the blue side of [Ar {\sc iv}]$\lambda$4740 is the blend of
the [Ne {\sc iv}]$\lambda \lambda$4724,4726 doublet (Figures
2b and 2c). This feature may contribute to the red-side tail of the
[Ar {\sc iv}]$\lambda$4712 emission. The emission-line feature seen 
at the blue side of [Fe {\sc vi}]$\lambda$5631
is identified as [Ca {\sc vii}]$\lambda$5616.
Since the wavelengths of the [Fe {\sc vii}]$\lambda$5721 and the
[N {\sc ii}]$\lambda$5755 emission are close in each other on the spectrum,
the red-side tail of the [Fe {\sc vii}]$\lambda$5721 emission and the
blue-side tail of the [N {\sc ii}]$\lambda$5755 are affected by their 
neighbor lines (particularly, the latter one is). To show how the 
[Fe {\sc vii}]$\lambda$5721 emission affects the [N {\sc ii}]$\lambda$5755 
profile, we give a larger portion of the spectral region around the 
[N {\sc ii}]$\lambda$5755 emission in Figure 3. The weak emission feature 
seen at the $\sim +700$ km s$^{-1}$ of the [O {\sc i}]$\lambda$6300 
emission is [S {\sc iii}]$\lambda$6312. We do not investigate these 
faint emission lines since the S/N of the data is not enough high to 
discuss their kinematic properties. Note that a possible contribution of 
[Ca {\sc v}]$\lambda$6087 to [Fe {\sc vii}]$\lambda$6087 has been 
sometimes suspected for spectra of Seyfert nuclei (e.g., Koski 1978; 
see also Appenzeller \& \"{O}streicher 1988). 
However, simple photoionization models predict that the flux ratio 
of [Ca {\sc v}]$\lambda$6087/[Fe {\sc vii}]$\lambda$6087 is always 
$\lesssim$ 0.1 for reasonable parameter ranges, which is confirmed 
by our calculations\footnote{
In the performed models, one-zone, dust-free gas clouds with the
elemental abundances of the solar ones are assumed.
The parameter ranges we examined are $n = 10^2 - 10^7$ cm$^{-3}$ 
and $U = 10^{-3.5} - 10^{-1.5}$ where $n$ and $U$ are the 
hydrogen density and the ionization parameter, respectively.
} by using the publicly available code $Cloudy$ version 94.00 
(Ferland 1997, 2000). We thus neglect the possible
contamination of the [Ca {\sc v}]$\lambda$6087 emission to the
[Fe {\sc vii}]$\lambda$6087 emission in the following discussion.

\begin{figure*}
\epsscale{0.6}
\plotone{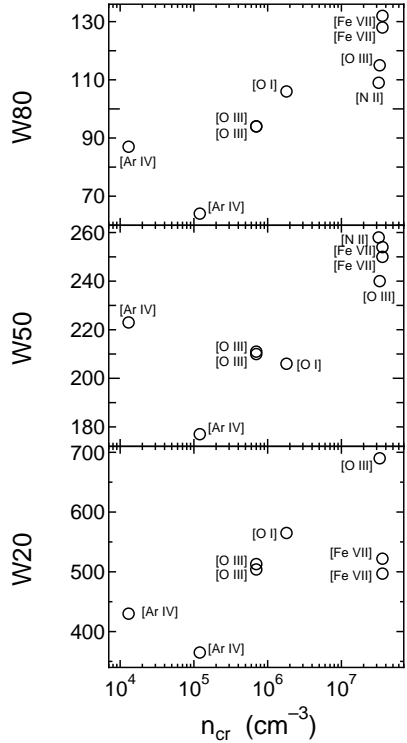}
\caption{
Measured line widths as a function of the critical density of the 
corresponding transitions. $W80$, $W50$, and $W20$ are presented 
in the upper, middle, and lower panels,
respectively. Note that the data of [O {\sc iii}]$\lambda$4959 and
[O {\sc iii}]$\lambda$5007 are indistinguishable in the upper panel.
\label{fig4}}
\end{figure*}

We quantify the observed profiles of the emission lines presented in
Figure 2, basically by adopting the manner of Heckman et al. (1981).
For each emission line, we measure the full line widths at levels
of 80\%, 50\%, and 20\% of the peak intensity above the continuum
($W80$, $W50$, and $W20$, respectively). Note again that the effects of the
instrumental broadening are not corrected owing to the high
wavelength resolution of our observation.
We also measure the velocities of the line centers at the three
respective levels ($C80$, $C50$, and $C20$). 
To quantify the asymmetries of each emission line, we measure
the widths between $C80$ and the edges of emission line of both side
($WL20$ and $WR20$ for the left- and the right-side width from $C80$;
see Figure 2 of Heckman et al. 1981).
And following to the manner of Heckman et al. (1981), we calculate an
asymmetry index $AI_{20} = (WL20 - WR20) / (WL20 + WR20)$.
Positive figures of this parameter
denote the existence of pronounced excess at the blue side of
the emission line, and vice versa. 
The measured parameters, i.e., $W80$, $W50$, $W20$, $C80$, $C50$, $C20$, and 
$AI_{20}$, for each emission line, are presented in Table 2. 
The measured recession velocities ($C80$, $C50$, and $C20$)
have once converted into the heliocentric ones, and the presented velocities
are relative values to the (heliocentric) systemic recession velocity 
of NGC 4151 determined by the H {\sc i} 21 cm emission 
($v_{\rm sys} = 995$ km s$^{-1}$; Heckman, Balick, \& Sullivan 1978).
We do not measure $W20$, $C20$, and $AI_{20}$
for [N {\sc ii}]$\lambda$5755, since the blue
wing of this line appears to be significantly affected by the red tail of the
[Fe {\sc vii}]$\lambda$5721 emission (see Figure 3).

It is apparently shown that the observed forbidden lines are generally
blueshifted relative to the systemic recession velocity of the host
galaxy, NGC 4151.
This has been frequently reported for many Seyfert galaxies
(e.g., Heckman et al. 1981; Penston et al. 1984; Whittle 1985a;
Schulz 1987; Veilleux 1991c).
The observed value of $C80$ for [O {\sc iii}]$\lambda$ of NGC 4151
is consistent with the previous high-dispersion spectroscopies
(--45 km s$^{-1}$, Vrtilek \& Carleton 1985; --40 km s$^{-1}$, Schulz 1987).
The tendency that the asymmetry indices of most of emission lines suggest
the blue-wing excess has been also seen for many Seyfert galaxies
(e.g., Heckman et al. 1981; De Robertis \& Osterbrock 1984;
Vrtilek \& Carleton 1985; Whittle 1985a; Veilleux 1991c; Moore et al. 1996).
The measured emission-line widths tend to be narrower than the ones
reported based on medium-dispersion (i.e., $\Delta v \gtrsim 100$ km s$^{-1}$)
spectroscopic observations of NGC 4151 (e.g., Heckman et al. 1981;
De Robertis \& Osterbrock 1984; Bochkarev, Shapovalova, \& Zhekov 1991).
This is thought to be the effect of the instrumental broadening.
As described by Veilleux (1991c), the effect of the instrumental broadening
on medium velocity-resolution spectra
cannot be corrected completely by the classical quadratic subtraction
method. Since the wavelength resolution of our observation is high
enough, our results on the emission-line widths are much
more reliable than those derived by the previous medium-dispersion 
spectroscopic observations.

\begin{figure*}
\epsscale{0.6}
\plotone{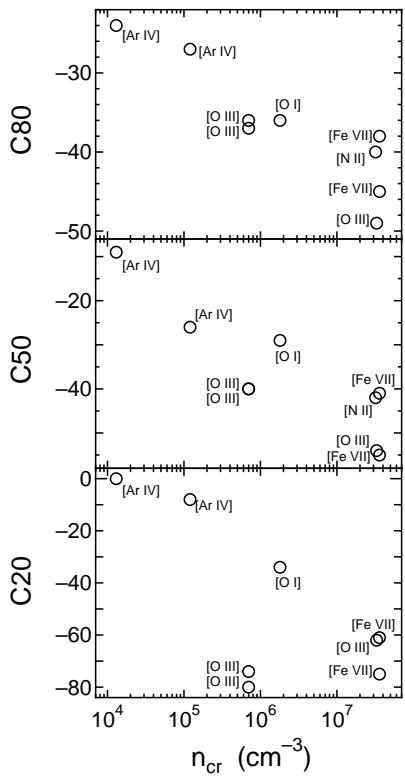}
\caption{
Measured central recession velocity as a function of the critical density 
of the corresponding transitions. $C80$, $C50$, and $C20$ are 
presented in the upper, middle, and lower panels,
respectively. Note that the data of [O {\sc iii}]$\lambda$4959 and
[O {\sc iii}]$\lambda$5007 are indistinguishable in the middle panel.
\label{fig5}}
\end{figure*}

We then investigate how the measured profile parameters of each line
depend on or independent of the critical density and the ionization
potential of the corresponding transitions.
The full width and the central recession velocity of
each level are plotted as a function of the critical density
(Figures 4, 5) and of the ionization potential (Figures 6,7).
In Figure 8, the asymmetry index $AI_{20}$ is plotted as functions
of the critical density and the ionization potential.
As shown these figures apparently, clear correlations are seen between
the measured profile parameters and the critical density while
such correlations are not seen between the profile parameters and the 
ionization potential. To be more concrete, emission lines with a
higher critical density show broader widths and bluer central recession
velocities at any intensity level.
In order to investigate the statistical significance of these tendencies,
we apply Spearman's rank-order correlation statistical test on the data.
The null hypothesis is that each measured profile parameter is not
correlated with the critical density or the ionization potential of
the corresponding emission line. The resultant probabilities are given
in Table 3. It is now statistically confirmed that any of the measured profile
parameters are not correlated with the ionization potential of the
corresponding ions while the parameters tend to correlate with the
critical density of the transitions. The statistical significances of
the correlations concerning the critical density are high for the core 
properties (i.e., $W80$ and $C80$) while those are statistically not 
significant for the wing properties (i.e., $W20$, $C20$, and $AI_{20}$).
The difference between the core and the wing may be intrinsic, but
we cannot exclude the possibility that the measured base properties
are affected by the blaze profile of the spectrum described in Section 2.4
and Appendix A. There is no statistically significant correlations between
the asymmetry index $AI_{20}$ and the critical density, and
between $AI_{20}$ and the ionization potential.

\begin{figure*}
\epsscale{0.6}
\plotone{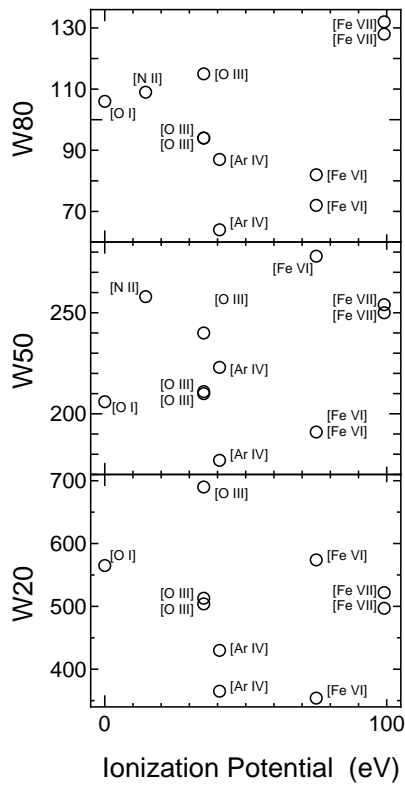}
\caption{
Measured line widths as a function of the ionization potential of the 
corresponding ions. $W80$, $W50$, and $W20$ are presented 
in the upper, middle, and lower panels,
respectively. Note that the data of [O {\sc iii}]$\lambda$4959 and
[O {\sc iii}]$\lambda$5007 are indistinguishable in the upper panel.
\label{fig6}}
\end{figure*}

\section{DISCUSSION}

As described in the last section, the emission-line spectrum of
NGC 4151 has the following two remarkable aspects.
(1) The widths of the forbidden emission lines are well correlated 
with the critical density of the transitions while they are not 
correlated with the
ionization potential of the corresponding ions.
(2) The blueshift amounts of the forbidden emission lines are also well
correlated with the critical density while there is no significant
correlation between the blueshift amount and the ionization potential.
We interpret these observational results and accordingly discuss the
geometrical and kinematical nature of the NLR in NGC 4151.
However, prior to discuss them, we have to mention the
effects of a possible aperture effect on these results. 
Since the lower-ionization and the lower-critical-density lines tend 
to arise at the further distant places from the nucleus in general, 
we may miss the narrow, symmetric components for low-ionization and/or
low-critical-density lines. Nevertheless the significant correlations
between the critical density and the emission-line profiles are found.
Since the possible aperture effect should work in an opposite trend,
we can conclude that the observed correlations between the critical 
density and the emission-line profiles are real.

 \subsection{Emission-Line Width}

Not only NGC 4151, many Seyfert galaxies also show similar tendencies
about the emission-line widths; i.e., line widths are correlated
with the critical density and/or the ionization potential
(e.g., Filippenko \& Halpern 1984; De Robertis \& Osterbrock 1984, 1986;
Veilleux 1991a; Alloin et al. 1992).
The most simple interpretation for this is that the gas clouds are
stratified in NLRs. In this scheme,
the gas clouds with larger gas densities and with
higher ionization degrees are located closer to the nucleus,
where the gravitational potential is deeper than the outer part of NLRs
and thus the velocity dispersion of gas clouds becomes relatively large.
However, this simple scheme may be difficult to explain the cases
that the emission-line widths are correlated only with 
the critical density or the ionization potential.

As for the case of NGC 4151, one of the high-critical-density forbidden 
emission lines, [N {\sc ii}]$\lambda$5755, has a broader velocity width 
than the other low-critical-density emission lines despite its 
low-ionization potential, as shown in Figures 2, 3, and 4. A velocity 
width is thought to be closely related with a depth of the gravitational 
potential at the effective emitting region of the corresponding emission 
line. Thus the broadness of the [N {\sc ii}]$\lambda$5755 emission 
suggests that emission lines with low-ionization potentials can arise at 
inner NLRs, at least in the nucleus of NGC 4151. How 
situation enables to realize such a condition? To investigate this 
issue, one of the most powerful approaches would be the locally 
optimally emitting cloud (LOC) photoionization model, which was 
originally constructed to explore the nature of broad-line regions 
in AGNs by Baldwin et al. (1995) and which was applied to NLRs later 
(Ferguson et al. 1997a, 1997b). In this model, gas clouds with a 
wide range of physical conditions are present at a wide range in the 
radial direction, and thus the net emission-line spectra can be 
calculated by integrating in the parameter space of the gas density 
and the radius, assuming some distribution functions. Therefore we 
can know the emissivity distribution of each emission line in the
radial direction by integrating the results of this LOC models only
for the gas density which can be used to discuss the difference
of emission-line widths of various forbidden emission lines, assuming
that the velocity dispersion of gas clouds is determined by the
gravitational potential caused by the central supermassive black hole
(see Ferguson et al. 1997a). Here we should mention that this 
assumption is rather inappropriate when gas clouds at outer NLRs are 
considered, because the gravitational potential caused by the host 
galaxy is not negligible at such an outer NLR. However, it seems 
very difficult to take the contribution of the host galaxy on the 
kinematics of NLR clouds into account adequately. This is because 
the gas motion of NLR clouds tends to be independent of the galactic 
rotation (e.g., Heckman et al. 1989), except for clouds at an outer
part of very extended NLRs (see Pedlar et al. 1992 for the outer 
NLR in NGC 4151). The widths of forbidden lines tend to be correlated 
with the velocity dispersion of the bulge of host galaxies (e.g.,
Nelson \& Whittle 1996; Boroson 2003), although such an effect is hard 
to be quantified for each point in NLRs. 
Thus, as for the gas clouds at outer NLRs, the 
moving velocities inferred by the LOC model tend to be smaller than 
the actual situation where the host galaxy contributes to the
gravitational potential. We should be aware of these matters in the
following discussion, although we do not take the effects of the
host galaxy on the gas kinematics into account in this paper.

\begin{figure*}
\epsscale{0.6}
\plotone{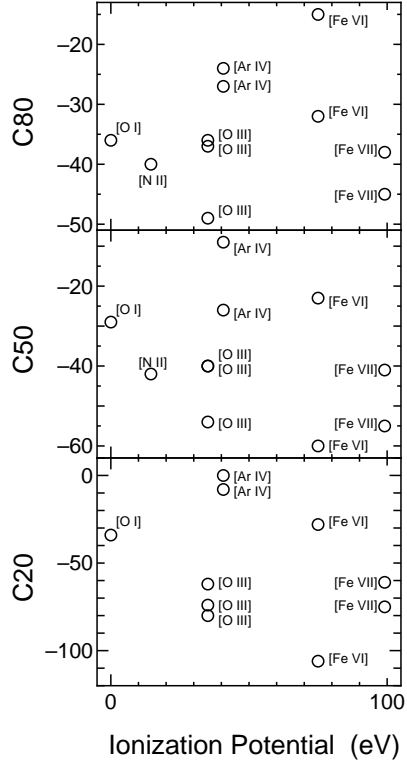}
\caption{
Measured central recession velocity as a function of the ionization
potentials of the corresponding ions. $C80$, $C50$, and $C20$ are 
presented in the upper, middle, and lower panels,
respectively. Note that the data of [O {\sc iii}]$\lambda$4959 and
[O {\sc iii}]$\lambda$5007 are indistinguishable in the middle panel.
\label{fig7}}
\end{figure*}

As shown in Figure 1 of Ferguson et al. (1997a),
the emission-line strength of each transition reaches its peak where the 
gas density is comparable to the critical density of the transition
and the ionization parameter (i.e., the radius) is preferred by
the ionization degree of the corresponding ion.
Therefore the radius of peak emissivity is in proportion to
the square root of the critical density for a given ionization potential,
which is apparent when the line-strength distributions of pairs of 
[O {\sc iii}]$\lambda \lambda$4363,5007, 
[S {\sc ii}]$\lambda \lambda$6717,6731, or
[O {\sc ii}]$\lambda \lambda$3727,7325 are compared in each other
(see Figure 1 of Ferguson et al. 1997a).
The radius is also in proportion to the square root of the ionization
parameter for a given gas density by the definition.
Here we focus on the behavior of the emission-line strengths 
of transitions with a high critical density but with a low ionization
potential, such as [N {\sc ii}]$\lambda$5755 and 
[O {\sc i}]$\lambda$6300, on the density-radius planes in Figure 1
of Ferguson et al. (1997a).
Such emission lines can arise from a cloud with a high density and
at far from the nucleus. However, the ionization parameter of such
a cloud becomes very low, and thus highly ionized ions cannot exist
in such circumstances. Affected by the existence of such high-density
and low-ionization-parameter gas clouds, the peak radius of 
emission lines with a high critical density and a low ionization potential
tends to shift outward. That is, the radius of emissivity peaks
depends more strongly on the ionization potential of the ions than
on the critical density of the transition.
See Appendix B where this issue is investigated more quantitatively.

De Robertis \& Osterbrock (1986) also discussed this issue; i.e.,
the emission-line widths would be mainly dependent with the ionization 
potential of the ions if the density distributions
are completely independent of the radius from the nucleus.
They discussed the effects of the radial distance dependences of the
gas density as follows. If the radial distance dependence of the
density can be simply described by a power-law form,
$n(r) \propto r^{-\xi}$, then the ionization parameter of a cloud is
written as
\begin{equation}
U(r) \propto \frac{L_{\rm ion}}{4 \pi r^2 n(r)} \propto r^{\xi - 2}.
\end{equation}
Therefore, for the case of $\xi = 0$, the ionization parameter
depends on the radius as $U \propto r^{-2}$ and thus 
the emission-line widths are expected to correlate only with 
the ionization potential but not to correlate with the critical density.
On the contrary, for the case of $\xi = 2$, the ionization parameter
becomes to be independent of the radius and thus the the 
emission-line widths are expected not to correlate with the 
ionization potential but to correlate with the critical density.
Therefore, to explain the observed properties of emission-line profiles of 
NGC 4151, we should consider the radial distance dependences of the
gas density.

\begin{figure*}
\epsscale{1.8}
\plotone{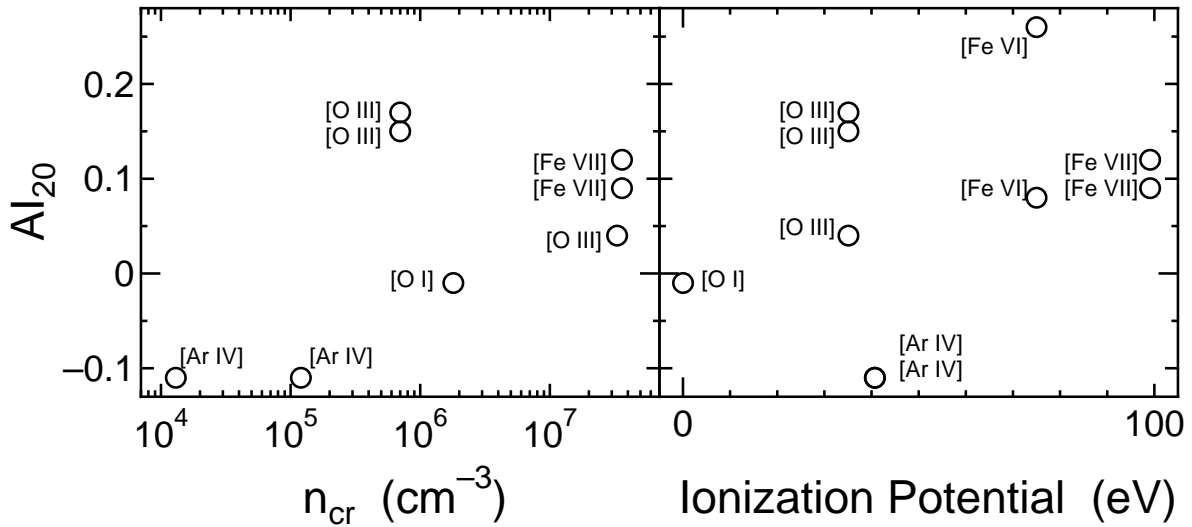}
\caption{
Measured asymmetry index, $AI_{20}$, as functions of the
critical density (left panel) and of the ionization potential
(right panel). Note that the data of [Ar {\sc iv}]$\lambda$4712 and
[Ar {\sc iv}]$\lambda$4740 are indistinguishable in the right panel.
\label{fig8}}
\end{figure*}

Here one important concern is
that NGC 4151 is a type 1.5 Seyfert galaxy and thus the inner regions
of the dusty torus are not obscured. 
It is known that type 1 AGNs exhibit statistically stronger
high-ionization forbidden emission lines in their spectra than
type 2 AGNs (e.g., Cohen 1983; Murayama \& Taniguchi 1998a; 
Nagao et al. 2000, 2001c).
This suggests that the high-ionization lines emitting regions
(hereafter ``high-ionization nuclear emission-line regions (HINERs)'';
Binette 1985; Murayama, Taniguchi, \& Iwasawa 1998)
are located very close to the nuclei (i.e., inner regions of dusty tori)
and thus can be hidden 
when we see the nucleus from an edge-on view toward the dusty tori
(e.g., Murayama \& Taniguchi 1998a; Barth et al. 1999; 
Nagao et al. 2000; Tran et al. 2000; Nagao et al. 2001c)
like broad-line regions.
Murayama \& Taniguchi (1998b) showed that this HINER component 
significantly contributes to the observed emission-line spectra of
NLRs in type 1 AGNs (see also Nagao et al. 2001b).
Therefore it seems natural to add this HINER component into the LOC model.
If the gas clouds in HINER consist of preferentially high-dense
ones, only high-critical-density emission lines are enhanced by
introducing the HINER component.
This assumption is consistent with the result of Nagao et al. (2001c) 
that the gas density of this HINER component should be very 
dense (i.e., $> 10^6$ cm$^{-3}$) to explain the observed 
emission-line flux ratios of high-ionization forbidden lines.
Taking this HINER component into account, the observed correlation
between the emission-line width and the critical density of the 
transitions can be understood.
See Appendix B where the effects of adding the HINER component into
the LOC model are investigated.

 \subsection{Emission-Line Shifts}

We then discuss the origin of the observed correlation between
the blueshift amounts of the emission-line center and the critical 
densities of the transition shown in Figure 5.
It has been often reported that
the observed forbidden lines in Seyfert galaxies are 
blueshifted relative to the systemic recession velocity of the host
galaxies (e.g., Heckman et al. 1981; Penston et al. 1984; 
Whittle 1985a; Schulz 1987; Veilleux 1991c).
This has been interpreted that the gas clouds flow radially with
some opacity sources, probably dust grains (e.g., Heckman et al. 1981).
That is, if gas clouds flow outward in the medium with dust grains,
outflowing clouds in the far side (i.e., in the receding hemisphere)
suffer heavier extinction and thus the net blueward shift of 
emission lines can occur. Alternatively, infalling gas motions may
be the origin of the blueshift if the gas clouds contain plenty of
internal dust grains, because such gas clouds are ionized only on
the side facing the ionizing continuum source and thus we can see
only the receding gas clouds in the far side from an observer.
Then, which situation is the case for NGC 4151? We investigate 
this problem by focusing on the observed correlation between 
blueshift amounts of the emission-line center and the critical 
densities of the transitions.

%------------------------------------------------------------------
%                Table 3
%------------------------------------------------------------------
\begin{deluxetable}{lcc}
\tablenum{3}
\tablecaption{Results of the Rank-Order Correlation Test\tablenotemark{a}}
\tablewidth{350pt}
\tablehead{
\colhead{Profile Parameter} &
\colhead{Critical Density} &
\colhead{Ionization Potential}
}
\startdata  
$W80$\dotfill       & 4.4$\times 10^{-6}$ & 9.5$\times 10^{-1}$ \nl
$W50$\dotfill       & 4.7$\times 10^{-2}$ & 4.9$\times 10^{-1}$ \nl
$W20$\dotfill       & 1.3$\times 10^{-1}$ & 3.9$\times 10^{-1}$ \nl
$C80$\dotfill       & 2.6$\times 10^{-3}$ & 6.9$\times 10^{-1}$ \nl
$C50$\dotfill       & 1.8$\times 10^{-3}$ & 6.0$\times 10^{-1}$ \nl
$C20$\dotfill       & 2.4$\times 10^{-1}$ & 8.7$\times 10^{-1}$ \nl
$AI_{20}$\dotfill   & 3.3$\times 10^{-1}$ & 5.8$\times 10^{-1}$
\enddata 
\tablenotetext{a}{Based on Spearman's rank-order correlation
                  statistical test. The presented probabilities are
                  the two-sided significance levels of the deviation
                  of the rank-order correlation coefficient
                  from the null hypothesis that the two parameters
                  are uncorrelated.}
\end{deluxetable}
%------------------------------------------------------------------

De Robertis \& Shaw (1990) examined the behaviors of emission-line
profiles for both the outflowing and infalling cases by performing
kinematical model calculations. They showed the following two 
important results. (1) As for the outflowing case, a concentration
of the dust distribution near the nucleus is required in order to
generate more blueward shifts of the lines arising at inner NLRs
than those arising at outer NLRs. (2) As for the infalling case,
on the other hand, more blueward shifts of the lines arising at
inner NLRs can be caused if high-dense gas clouds at such an inner
NLR contains much internal dust grains. However, neither the former 
nor the latter situation seems unrealistic, because Nagao et al. 
(2003) recently reported that iron is not depleted in gas clouds in
HINERs and thus the innermost part of NLRs is thought to be a
dust-free circumstance. Here we it should be mentioned that 
De Robertis \& Shaw (1990) pointed out that the existence of a dusty
torus around the nucleus may play an important role when the 
profiles of high-ionization emission lines are discussed since they
arise at the innermost part of NLRs, whose spatial extension may
comparable to the scale of dusty tori. Taking this point into account,
we propose the following situation to explain the observation.
That is, in HINERs, the gas clouds without dust grains flow outward 
in the dust-free circumstance. However, due to the existence of the 
dusty torus, the receding gas clouds are obscured from an observer.
The observed correlation between the blueshift amounts of the emission 
lines and the critical densities is then explained if the gas clouds 
located closer to the nucleus have higher gas densities and larger 
outflowing velocities. The schematic view of this idea is shown in 
Figure 9. Note that Erkens, Appenzeller, \& Wagner (1997) also 
mentioned that the high-ionization-line emitting clouds flow outward 
radially, which causes relatively broad widths and blueward velocity 
shifts of high-ionization forbidden emission lines.

It should be noted that the above model shown in Figure 9 requires
a special viewing angle from the observer toward the nucleus, since
the receding HINER component is hard to be obscured by the torus
when seen from a pole-on view toward the torus. The receding clouds
are hidden effectively when our line of sight is near the edge of
the torus. Indeed, there are a number of evidence which suggests that
the viewing angle toward the nucleus of NGC 4151 is just to be
such a special one. One of them is the type-switching phenomenon
seen in NGC 4151 (e.g., Ayani \& Maehara 1991); i.e., the broad
component of Balmer lines has been disappeared from the optical
spectrum of NGC 4151 (see also Antonucci \& Cohen 1983; Penston \&
Perez 1984). This AGN-type switching phenomenon has been sometimes
thought to be caused by a time variation of the optical thickness of
patchy tori (e.g., Goodrich 1989; Tran, Osterbrock, \& Martel 1992;
Loska, Czerny, \& Szczerba 1993; Aretxega et al. 1999). It has been
pointed out that this variation results from the special viewing angle 
(i.e., through near the edge of the torus), which is consistent with the 
model proposed here (see, e.g., Kriss et al. 1995; Weymann et al. 1997). 
The column density of the neutral hydrogen in the line of sight toward 
the nucleus of NGC 4151 is also remarkable since it is significantly 
large among type 1 and type 1.5 Seyfert galaxies ($N_{\rm H} \sim 10^23$ 
cm$^-2$; e.g., Maisack \& Yaqoob 1991; Yaqoob \& Warwick 1991; 
Zdziarski, Johnson, \& Magdziarz 1996; Warwick et al. 1996; 
George et al. 1998; Yang, Wilson, \& Ferruit 2001). 
It is remarkable that the X-ray spectra taken by $ASCA$ should consist
of both absorbed and non-attenuated radiation (George et al. 1998).
This implies that the absorbing material in the line of sight toward
the nucleus has patchy structure and thus only a part of the nuclear 
emission can reach up to us without the attenuation. It should be
also noted that the column density of the neutral hydrogen
toward the nucleus of NGC 4151 is temporarily variable significantly
(Yaqoob, Warwick, \& Pounds 1989; Warwick et al. 1989;
Fiore, Perola, \& Romano 1990; Yaqoob \& Warwick 1991).
Recent high-spatial resolution optical spectroscopic observations 
have also suggested that the nucleus of NGC 4151 is viewed from 
the viewing angle near the edge of the torus (see, e.g., Figure 7
of Hutchings et al. 1998). All of the above facts suggests that
we see the nucleus of NGC 4151 from the special viewing angle near 
the edge of the torus and thus the receding gas clouds in HINERs
is effectively obscured by the torus, which is completely consistent 
with the model proposed for NGC 4151 in this paper.

\begin{figure*}
\epsscale{1.2}
\plotone{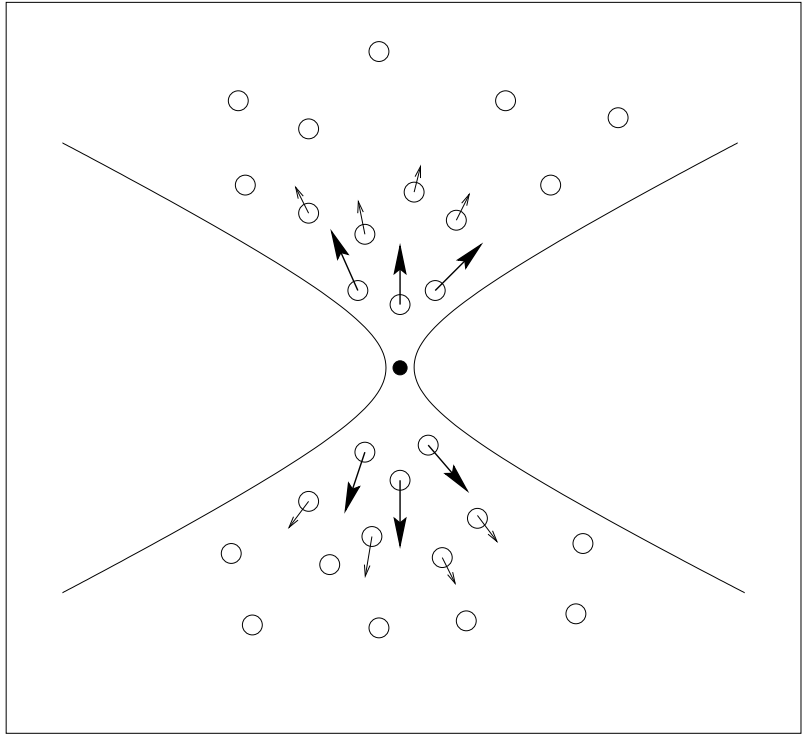}
\caption{
Schematic view of the inner region of the NLR in NGC 4151
inferred from our study.
The gas clouds located closer to the nucleus in the NLR
have higher gas densities and larger outflowing velocities.
The receding gas clouds are obscured from an observer by the dusty torus.
The filled circle denotes the broad-line region and the central engine
of NGC 4151.
\label{fig9}}
\end{figure*}

\section{SUMMARY}

We carried out high-dispersion spectroscopy ($R \sim 45,000$) of 
the NLR in NGC 4151 by using Subaru/HDS.
The main findings by examining the obtained emission-line spectrum
are as follows.

\begin{itemize}
  \item The emission-line centers are generally blueshifted
        relative to the systemic recession velocity of the host galaxy.
  \item The widths and the blueshift amounts of emission lines are 
        well correlated with the critical densities of the transitions.
  \item The widths and the blueshift amounts of emission lines show no
        significant correlations with the ionization potentials of the ions.
\end{itemize}

To interpret the origin of these properties, 
we refer the results of the photoionization model 
calculations performed by Ferguson et al. (1997a), i.e., the LOC model.
It is suggested that simple power-law distribution of
gas density which is independent of the radius from the nucleus cannot
explain the observed correlation between the emission-line widths and
the critical densities of the transitions.
However, taking the dense gas clouds expected to exist in HINERs
into account, high-critical-density
emission lines can arise at relatively inner regions of NLRs even if
their ionization potential is low. We note that
the observed correlation between the blueshift amounts of emission lines
is also explained if such dense gas clouds located closer 
to the nucleus have larger outflowing velocities.

%------------------------------------------------------------------
%                Table 4
%------------------------------------------------------------------
\begin{deluxetable}{lcc}
\tablenum{4}
\tablecaption{Conditions for the Maximum Equivalent Width of Emission Lines}
\tablewidth{360pt}
\tablehead{
\colhead{Emission Line} &
\colhead{Density} &
\colhead{Radius \tablenotemark{a}} \\
\colhead{} &
\colhead{(cm$^{-3}$)} &
\colhead{(cm)} 
}
\startdata  
[O {\sc i}]$\lambda$6300\dotfill    & 10$^{8.0}$ \tablenotemark{b} 
                                                 & 10$^{21.2}$ \nl 
[Ar {\sc iv}]$\lambda$4712\dotfill  & 10$^{3.5}$ & 10$^{19.7}$ \nl 
[Ar {\sc iv}]$\lambda$4740\dotfill  & 10$^{5.0}$ & 10$^{18.7}$ \nl 
[O {\sc iii}]$\lambda$5007\dotfill  & 10$^{4.7}$ & 10$^{19.2}$ \nl 
[O {\sc iii}]$\lambda$4363\dotfill  & 10$^{6.8}$ & 10$^{18.3}$ \nl 
[N {\sc ii}]$\lambda$5755\dotfill   & 10$^{6.3}$ & 10$^{19.7}$ \nl 
[Fe {\sc vii}]$\lambda$6087\dotfill & 10$^{6.1}$ & 10$^{18.7}$
\enddata 
\tablenotetext{a}{For the case of $L_{\rm ion} = 10^{43.5}$ ergs s$^{-1}$.
                  This scales with the luminosity as $L_{\rm ion}^{0.5}$.}
\tablenotetext{b}{This is a lower limited value since our calculations
                  examine the line emissivities only in the range of 
                  $n = 10^{1.0 - 8.0}$ cm$^{-3}$. See text for this issue.}
\end{deluxetable}
%------------------------------------------------------------------

\acknowledgments

We thank W. Aoki, A. Tajitsu, and S. S. Fujita for assisting 
our observation. We are grateful to all the staffs of the 
Subaru telescope, especially to the HDS instrument team.
We acknowledge P. A. M. van Hoof for giving us useful comments on
the rest-frame wavelengths of weak forbidden transitions, and
the anonymous referee for useful suggestions. 
We also thank G. Ferland for providing his code $Cloudy$ to the public. 
TN acknowledges financial support from the Japan Society for the
Promotion of Science (JSPS) through JSPS Research Fellowship for Young
Scientists. A part of this work was financially supported by Grants-in-Aid
for the Scientific Research (10044052, 10304013, and 13740122) of the 
Japanese Ministry of Education, Culture, Sports, Science, and Technology.

\appendix

\section{{\bf A.}\ \ \ BLAZE PROFILE CORRECTION}

\begin{figure*}
\epsscale{0.7}
\plotone{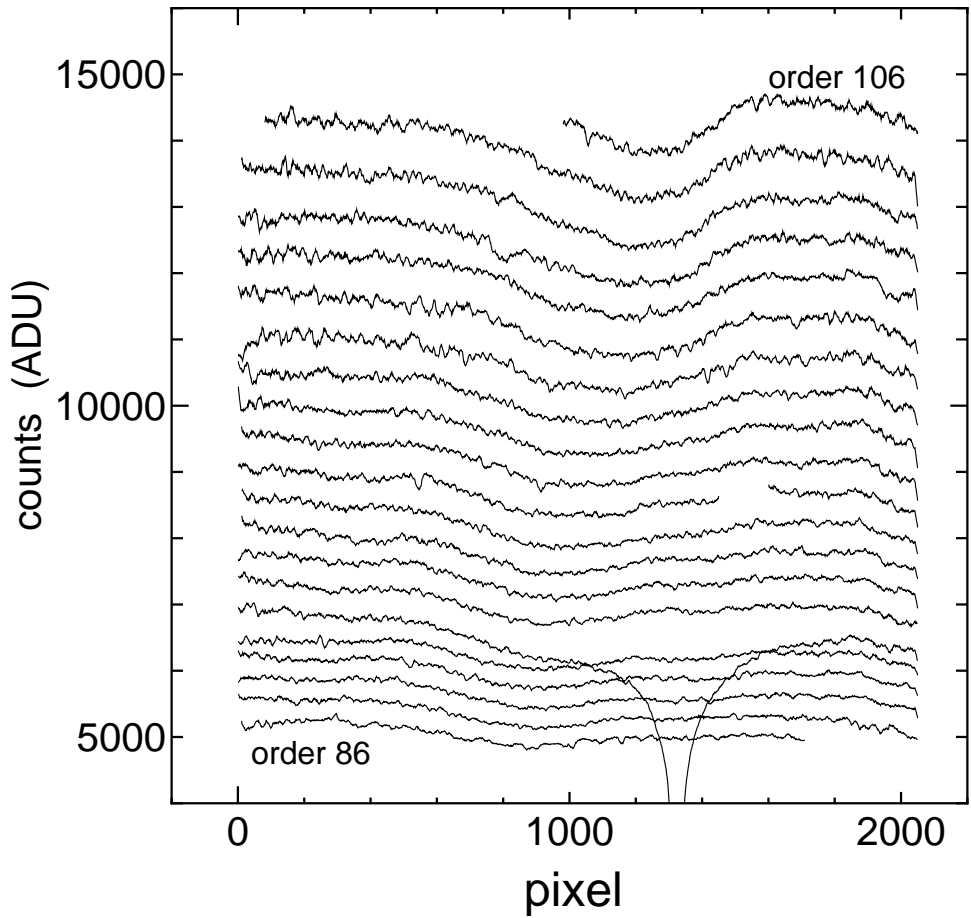}
\caption{
Spectrum of BD+26$\arcdeg$ 2606 before the blaze-function correction,
the wavelength calibration, and the flux calibration. The atmospheric
absorption features and the intrinsic metallic absorption lines are
removed. A 15-pixel smoothing is adopted. Only the spectra covered by
the red CCD, i.e., from the 86th order to the 106th order, are shown in
this figure.
\label{fig10}}
\end{figure*}

In this Appendix, we describe our method of the blaze profile
correction in detail (see also Section 2.4). In Figure 10, we give 
the spectra of the standard star BD+26$\arcdeg$ 2606 which are not 
flux-calibrated and are not corrected for the blaze profile. 
The atmospheric absorption features and intrinsic metallic absorption 
lines are removed from the spectra, and a 15-pixel smoothing is
performed. In this figure, the spectra have a curvature in some level, 
though they are expected to be intrinsically linear except for the 
region around a Balmer absorption feature (recall that 
BD+26$\arcdeg$ 2606 is a typical A-type star). This is just the effect of 
the unstability of the blaze profile pattern, because the blaze profile
pattern should be canceled when the standard-star frame has been 
divided by the flat flame, if the blaze profile pattern does not change. 
Although the blaze profile patterns are different among the orders, the 
difference is very gradual as a function of the echelle order. Since we 
know the intrinsic SED of BD+26$\arcdeg$ 2606, we can estimate the 
correction function for the remaining blaze profile patterns and can 
perform the blaze profile correction, only as for BD+26$\arcdeg$ 2606 itself. 
However, the estimated correction function about this standard star
cannot be used to carry out the blaze profile correction for NGC 4151, 
because the blaze profile is not the same between the standard-star 
frame and the object frame. Therefore we should make the blaze 
profile correction for NGC 4151 by using the spectrum of NGC 4151 
itself.

Being different from A-type stars, spectra of AGNs are far from 
featureless; i.e., various emission and absorption features are 
exhibited. However, as a first-order approximation, the power-law 
continuum emission component can be regarded to be linear in the 
wavelength range covered by one echelle order ($\sim 50 - 80 {\rm \AA}$).
Therefore, in the same way as for BD+26$\arcdeg$ 2606, the blaze profile of
the spectrum of NGC 4151 can be corrected by using the power-law
component of NGC 4151 itself, in principle. However unfortunately, the 
emission-line widths in the spectra of AGNs are so broad that we cannot 
extract the power-law component of NGC 4151 easily, except for some 
spectral regions without strong emission/absorption features. There are 
some roughly featureless wavelength regions in the spectrum of NGC 4151 
--- around 4600${\rm \AA}$, 5400${\rm \AA}$, and 6000${\rm \AA}$, where
the correction function for the blaze profile patterns can be estimated. 
Since the blaze profile pattern depends on the echelle 
order very smoothly (see Figure 10), we can perform the blaze profile 
correction for the whole range of the spectrum of NGC 4151 by using
the three estimated correction functions. Note that there are weak
emission/absorption features in the above three spectral regions.
Such features should be removed carefully by comparing the neighboring
echelle orders, when the correction functions are estimated.

As for the spectrum covered by the red CCD (i.e., from the 86th order
to the 106th order), we used the correction function estimated at
$\lambda \sim 6000 {\rm \AA}$ (the 99th and the 100th order). As for the
spectrum covered by the blue CCD (i.e., from the 108th to the 141st order),
we used the two correction functions, estimated at $\lambda \sim 4600 
{\rm \AA}$ (from the 128th to the 134th order) and at $\lambda \sim 5400
{\rm \AA}$ (from the 108th to the 114 th order). The former correction 
function is adopted for correcting the blaze profile of the echelle 
orders from the 108th to the 114th, while the latter one is adopted for 
the orders from the 128th to the 141st. As for the orders from the 115th 
to the 127th, the interpolated correction functions, which are made by 
using the above two correction functions, are used. The resultant 
blaze-profile-corrected spectrum of NGC 4151 are shown in Figure 1.

\section{{\bf B.}\ \ \ PHOTOIONIZATION MODEL CALCULATIONS}

\begin{figure*}
\epsscale{0.7}
\plotone{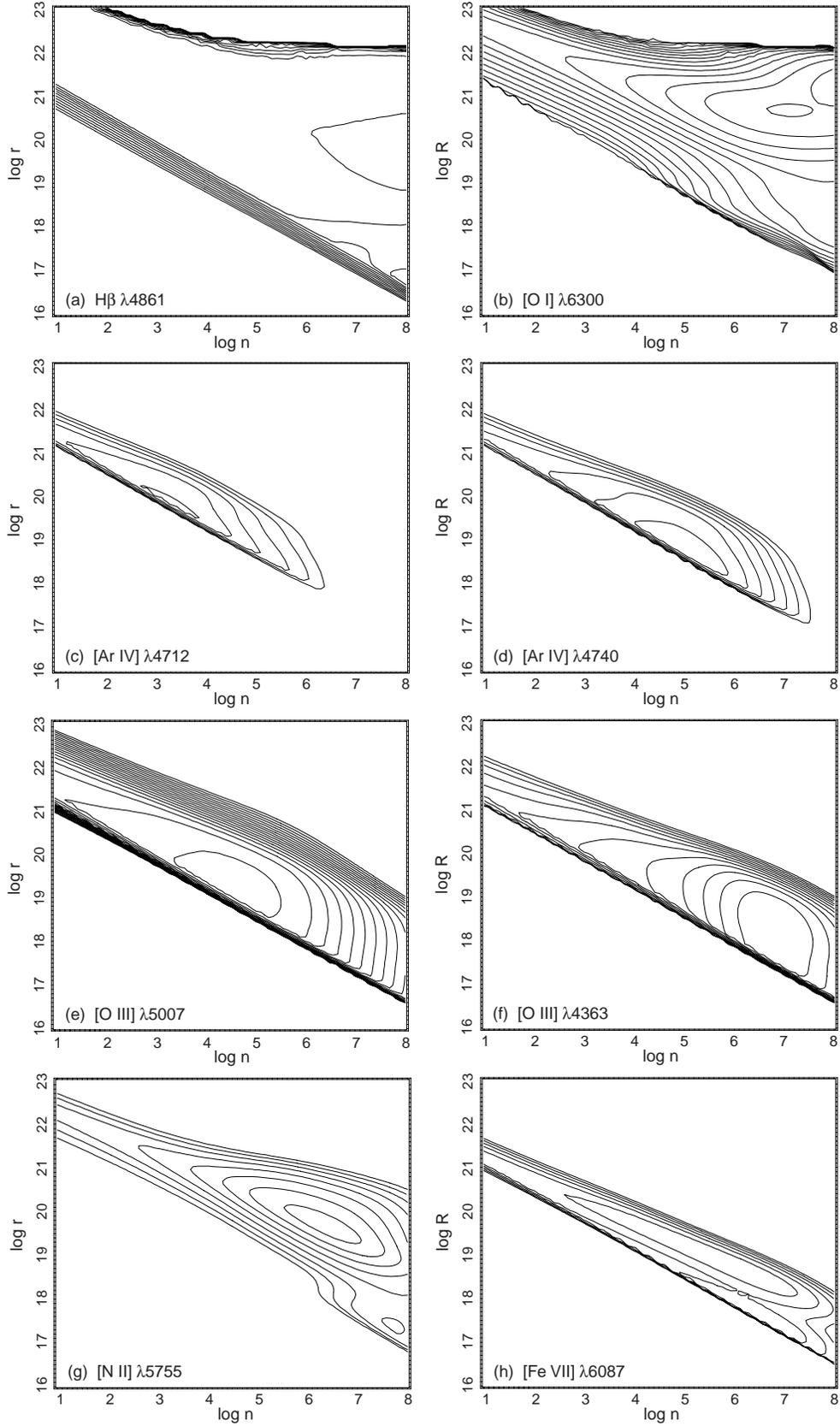}
\caption{
Contour maps of logarithmic equivalent widths of 8 emission lines
referred to the incident continuum at 4860${\rm \AA}$.
The lines are drawn with 0.2 dex steps down to the outer value of
1 ${\rm \AA}$.
(a) H$\beta \lambda$4861,
(b) [O {\sc i}]$\lambda$6300,
(c) [Ar {\sc iv}]$\lambda$4712,
(d) [Ar {\sc iv}]$\lambda$4740,
(e) [O {\sc iii}]$\lambda$5007,
(f) [O {\sc iii}]$\lambda$4363,
(g) [N {\sc ii}]$\lambda$5755, and
(h) [Fe {\sc vii}]$\lambda$6087.
\label{fig11}}
\end{figure*}

\begin{figure*}
\epsscale{0.51}
\plotone{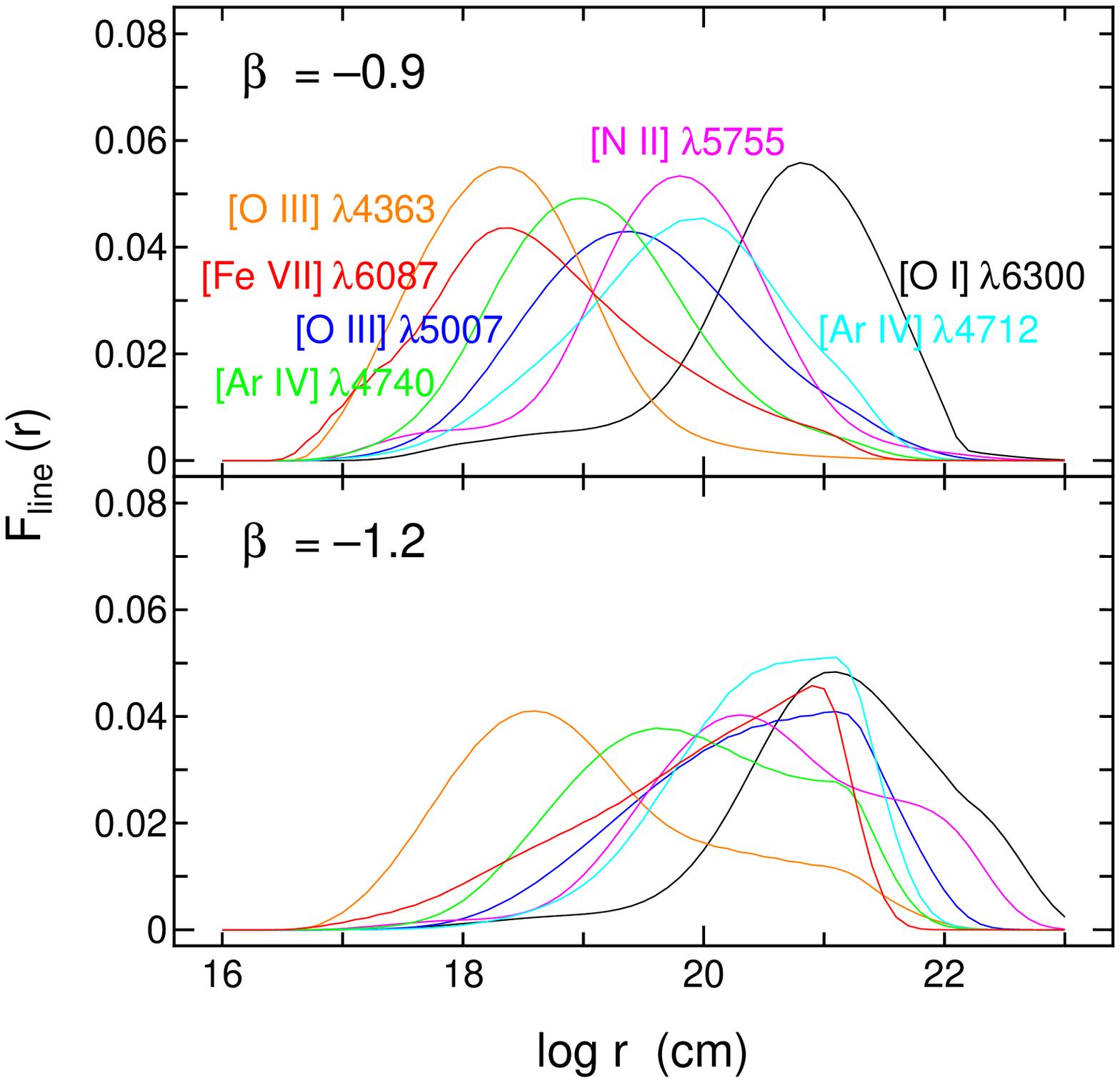}
\caption{
Radial distribution of emission-line fluxes of the forbidden lines
calculated by equation (2). The results for the two cases of the 
density integration are presented; $\beta = -0.9$ (upper panel) and
$\beta = -1.2$ (lower panel).
The scale of the vertical axis is arbitrary.
Note that the radius scales with the adopted luminosity of the 
ionizing continuum radiation as $L_{\rm ion}^{0.5}$.
The black, light blue, green, dark blue, orange, magenta, and red lines
denote the emission-line flux as a function of the radius, 
$F_{\rm line} (r)$, for
[O {\sc i}]$\lambda$6300, [Ar {\sc iv}]$\lambda$4712,
[Ar {\sc iv}]$\lambda$4740, [O {\sc iii}]$\lambda$5007,
[O {\sc iii}]$\lambda$4363, [N {\sc ii}]$\lambda$5755, and
[Fe {\sc vii}]$\lambda$6087, respectively.
\label{fig12}}
\end{figure*}

Here we present the results of our calculations
of the LOC photoionization model. This presentation mainly aims to
show the behaviors of emissivities of weak forbidden transitions
such as [Ar {\sc iv}]$\lambda \lambda$4712,4740 and 
[N {\sc ii}]$\lambda$5755 on the LOC model which were not presented
by Ferguson et al. (1997a), and to visualize
the effect of taking the contribution of the additional HINER
component into account.

For simplification, we assume dust-free, plane-parallel, and constant
density gas clouds with the chemical abundances of the solar ones,
which are taken from 
Grevesse \& Anders (1989) with extensions by Grevesse \& Noels (1993).
As for the spectral energy distribution (SED) of the input ionizing
continuum, we adopt the empirically constructed SED for typical
Seyfert galaxies (Nagao, Murayama, \& Taniguchi 2001a).
This SED is described by the following function;
\begin{equation}
f_{\nu} = \nu^{\alpha_{{\rm uv}}} \exp(-\frac{h\nu}{kT_{{\rm BB}}}) \exp
(-\frac{kT_{{\rm IR}}}{h\nu}) + a\nu^{\alpha_{{\rm x}}}
\end{equation}
(see Ferland 1997; Nagao et al. 2001a).
Here the following parameter set is adopted (Nagao et al. 2001a):
(i) the infrared cutoff of the big blue bump component, $kT_{\rm IR}$ = 
  0.01 Ryd,
(ii) the slope of the low-energy side of the big blue bump, 
  $\alpha_{\rm uv}$ = --0.5,
(iii) the UV--to--X-ray spectral index, $\alpha_{\rm ox}$ = --1.35,
(iv) the slope of the X-ray power-law continuum, $\alpha_{\rm x}$ = 
  --0.85, and
(v) the characteristic temperature of the big blue bump, $T_{\rm BB}$
  = 490,000 K.
Note that the parameter $a$ in the equation (1) is determined from
the adopted value of $\alpha_{\rm ox}$.
The last term in equation (1) is not extrapolated below 1.36 eV or
above 100 keV. Below 1.36 eV the last term is simply set to zero.
Above 100 keV the continuum is assumed to fall off as $\nu^{-3}$.
The luminosity of the ionizing continuum radiation is set to 
$L_{\rm ion} = 10^{43.5}$ ergs s$^{-1}$, which is a typical value 
for Seyfert galaxies. We investigate the output emission-line spectra
from a gas cloud with a hydrogen density of 
$n_{\rm H} = 10^{1-8}$ cm$^{-3}$ and a distance from the ionizing 
radiation of $r = 10^{16-23}$ cm. 
The ionization parameters of gas clouds are determined when 
a pair of hydrogen density and distance from the nucleus is given.
Note that this distance range
scales with the luminosity as $L_{\rm ion}^{0.5}$ and thus
does not mean an absolute one.
The calculation are stopped when one of the following three
conditions are met.
First is when the electron temperature falls to 3000 K, below which
the gas does not contribute significantly to the observed optical 
emission-line spectra. Second is when the cloud thickness exceeds
10\% of its distance from the central source in order to keep
gas clouds on the plane-parallel condition.
And third is when the column density of gas clouds exceeds 
10$^{24}$ cm$^{-2}$, beyond which the gas cloud becomes Thomson thick.
See Ferguson et al. (1997a, 1997b) for details of the stopping criteria.

We show the results of photoionization model calculations in Figure 11,
in which the contour maps of logarithm equivalent widths referred to
the incident continuum at 4860${\rm \AA}$ are presented.
The contours of calculated equivalent widths of the lines appear
to be roughly parallel to lines with a slope of
$d$ log $r$ / $d$ log $n$ = --0.5,
which corresponds to the condition of a constant ionization parameter.
Our results are almost the same as those of Ferguson et al. (1997a)
with small differences, which do not affect the following discussion
significantly. Those differences are thought to be due to 
the difference in the adopted shape of the input SED and in the
version of the code.

In Table 4, we give the pair of a hydrogen density and a radius from 
the nucleus which produces maximum equivalent width of each emission line.
Here we do not take the radial distribution function of gas clouds into
account, i.e., the values given in Table 4 mean the peak loci in the
contour maps in Figure 11.
Note that the density condition for the maximum equivalent width of
[O {\sc i}]$\lambda$6300, $n = 10^{8.0}$ cm$^{-3}$, is a lower limit value,
since our calculations examine the line emissivities only in the range of 
$n = 10^{1.0 - 8.0}$ cm$^{-3}$. By the calculations of 
Ferguson et al. (1997a), it is shown that the hydrogen density 
which produces the maximum equivalent width of [O {\sc i}]$\lambda$6300 
is $n \sim 10^{9}$ cm$^{-1}$ (see Figure 1 of Ferguson et al. 1997a).
As clearly presented in Figure 11 and Table 4,
the radius which produces the maximum equivalent width of each emission
line does not correspond to the critical density of each transition.
For instance, the radii of the maximum equivalent widths of
[O {\sc i}]$\lambda$6300 and [N {\sc ii}]$\lambda$5755 are
larger than those of [Ar {\sc iv}]$\lambda$4740 and 
[O {\sc iii}]$\lambda$5007.
However, this does not necessarily suggest that the effective 
line-emitting radius for emission lines is independent of the critical
density of the transitions.
This is because the effective line-emitting radius for each emission line
is determined by how the gas clouds distribute in the density.

In order to examine the radial distribution of the strengths of
the forbidden emission lines, the calculated emission-line fluxes should
be integrated taking the distribution function of the gas clouds in
density into account. The radial distribution of emission-line strength
is thus given by
\begin{equation}
F_{\rm line} (r) \propto \int F(r,n) f(n) dn
\end{equation}
where $F(r,n)$ is the emission-line flux of a single cloud with
a radius $r$ and a density $n$, and $f(n)$ is the adopted distribution
function of gas clouds on the density.

\begin{figure*}
\epsscale{0.51}
\plotone{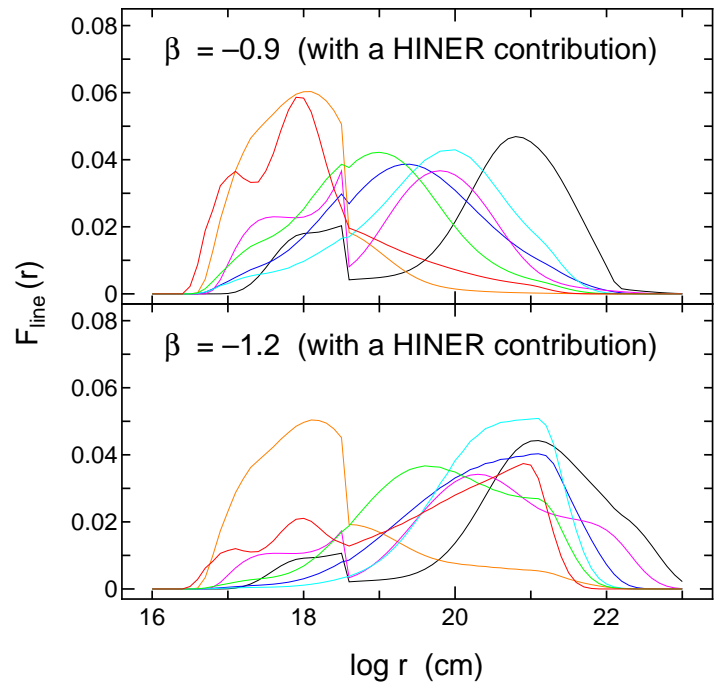}
\caption{
Same as Figure 12 but the contribution of
the HINER component is taken into account to the density integrations.
\label{fig13}}
\end{figure*}

First, for the most simplified case, we adopt a simple power-law form
as the distribution function of gas clouds on the density, i.e.,
$f(n) \propto n^{\beta}$. This is the assumption adopted by
Ferguson et al. (1997a) (see also Baldwin et al. 1995; 
Ferguson et al. 1997b). The density range of the integration is
from $n=10^{1.0}$ cm$^{-3}$ to $n=10^{8.0}$ cm$^{-3}$.
Note that Ferguson et al. (1997a) reported that there is no significant
difference on the results of integrations if gas clouds with 
$n > 10^{8.0}$ cm$^{-3}$ are included.
In Figure 12, we show the results of the density integrations of 
the emission-line fluxes in the cases of $\beta$ = --0.9 and --1.2.
In both cases, the effectively line-emitting radius of the emission lines
is mainly dependent with the ionization potential of the ions
but is not well dependent with the critical density of the transitions;
e.g., the effectively line-emitting radii of the
[O {\sc i}]$\lambda$6300 emission and the [N {\sc ii}]$\lambda$5755 
emission is relatively large comparing to other emission lines.
That is, the resultant radial dependences of emission lines
are not to be expected for NGC 4151.

Then, we consider the density integration taking the contribution of
the HINER component into account.
We adopt the following function
as a density distribution;
\begin{equation}
\cases{
       r > r_{\rm HINER} : f(n) = c n^{\beta}  \cr
       r \leq r_{\rm HINER} : f(n) = 
       \cases{c n^{\beta} \ \ {\rm for} \ \ n < n_{\rm HINER} \cr
             2 \times c n_{\rm HINER}^{\beta} 
                          \ \ {\rm for} \ \ n \geq n_{\rm HINER}
             \cr}
       \cr}
\end{equation}
where $c$ is an arbitrary constant, and the two parameters,
$r_{\rm HINER}$ and $n_{\rm HINER}$, characterize the HINER component
in the LOC model. Again the density range for the integration is
from $n = 10^{1.0}$ cm$^{-3}$ to $n = 10^{8.0}$ cm$^{-3}$.
We perform the density integration by equation (2) assuming
$r_{\rm HINER} = 10^{18.5}$ cm and $n_{\rm HINER} = 10^{7.0}$ cm$^{-3}$.
In Figure 13, we show the results of the density integrations of
the emission-line fluxes in the cases of $\beta$ = --0.9 and --1.2.
As shown in this figure clearly, emission lines with high critical
densities are enhanced in the HINERs while emission lines with
low critical densities are scarcely affected by introducing the 
HINER component. In other words, the dependences of emission-line
widths on the critical density of the transitions can be explained
by introducing the HINER component.

\clearpage
%-----------------------------------------------------------------------------

\end{document}